\documentclass[showpacs,aps,pra,superscriptaddress,twocolumn]{revtex4}
\usepackage{amsfonts,hyperref}
\usepackage{graphicx}
\usepackage{amssymb}
\usepackage{epsfig}
\usepackage{amsmath}
\usepackage{xcolor}
\begin{document}
\title{Gap Solitons and Bloch Waves in Nonlinear Periodic Systems}

\author{Yongping Zhang}
\affiliation{Institute of Physics, Chinese Academy of Sciences,
Beijing 100190, China}
\author{Zhaoxin Liang}
\affiliation{Shenyang National Laboratory for Materials Science,
Institute of Metal Research, Chinese Academy of Sciences, Wenhua
Road 72, Shenyang 110016, China}
\author{Biao Wu}
\affiliation{Institute of Physics, Chinese Academy of Sciences,
Beijing 100190, China}

\begin{abstract}
We comprehensively investigate gap solitons and Bloch waves in
one-dimensional nonlinear periodic systems. Our results show that
there exists a composition relation between them: Bloch waves at
either the center or edge of the Brillouin zone are infinite chains
composed of fundamental gap solitons(FGSs).  We argue that such a
relation is related to the exact relation between nonlinear Bloch
waves and nonlinear Wannier functions. With this composition
relation, many conclusions can be drawn for gap solitons without any
computation. For example, for the defocusing nonlinearity, there are
$n$ families of FGS in the $n$th linear Bloch band gap; for the
focusing case, there are infinite number of families of FGSs in the
semi-infinite gap and other gaps. In addition, the stability of gap
solitons is analyzed.  In literature there are numerical results
showing that some FGSs have cutoffs on propagation constant (or
chemical potential), i.e. these FGSs do not exist for all values of
propagation constant (or chemical potential) in the linear band gap.
We offer an explanation for this cutoff.



\end{abstract}
\date{\today}

\maketitle

\section{introduction}
With the state-of-the-art technology, various nonlinear periodic
systems have been experimentally realized \cite{Lederer,Morsch}.
Typical examples include nonlinear waveguide arrays
\cite{Christodoulides,Eisenberg}, optically induced photonic
lattices \cite{Fleischer}, and Bose-Einstein condensates (BECs) in
optical lattices \cite{Morsch,Brazhnyi}. In these nonlinear periodic
systems, there exist two typical stationary solutions,  Bloch waves
and gap solitons, which are keys to the understanding of such
systems.

Bloch wave is intrinsic to  periodic systems \cite{Aschcoft}. Hence
the concept of Bloch wave originally introduced for linear periodic
systems can be straightforwardly extended to the nonlinear periodic
systems. In both cases, Bloch waves are extensive and spread over
the whole space \cite{Wu1}. Nonlinearity, however, will
significantly affect the stability of Bloch waves. Whence, the
analysis of the stability of nonlinear Bloch waves (NBWs) has been a
focus of extensive research. For example, the nonlinearity-induced
instabilities of Bloch waves are directly responsible for the
formation of the train of localized filaments observed in various
optical systems \cite{Lederer,Meier2,Stepic}. For another instance,
the instability of Bloch waves has been experimentally observed for
BECs in optical lattices\cite{Burger,Fallani}, where the instability
is closely related to the breakdown of superfluidity in such
systems\cite{Wu1,Smerzi,Machholm1}.

In contrast, gap solitons are localized in space and only exist in
nonlinear periodic systems \cite{Lederer,Christodoulides}. So far
they have been found to exist in systems of different natures,
including nonlinear optical systems
\cite{Lederer,Xu,Sukhorukov,Cohen,Meier}, BEC systems
\cite{Louis,Efremidis,Pelinovsky}, even a surface system
\cite{Kartashov1}. One particularly important type of gap solitons
is fundamental gap solitons (FGSs), whose main peaks are located
inside a unit cell \cite{Louis,Efremidis,Mayteevarunyoo}. These FGSs
can be viewed as building blocks for the higher order gap solitons
\cite{Kartashov}. We note that other types of localized solutions
may also exist for nonlinear periodic systems, such as gap vortex in
two-dimensional nonlinear periodic systems \cite{Yang}. These
localized solutions persist in discrete models, where they are
called discrete soliton \cite{Lederer,Christodoulides} and discrete
vortex \cite{Malomed}, respectively.

A viewpoint has been floating in the community that the  NBWs (or
higher order gap solitons) can be regarded as chains composed of
FGSs in one-dimensional nonlinear periodic systems
\cite{Alexander,Carr,Kartashov,Smirnov}. Such a viewpoint would
occur to anyone who has observed the almost perfect match between a
NBW and the corresponding FGS as shown in Fig. \ref{First1}. In
particular, Alexander {\it et al.} \cite{Alexander} have found  that
a new set of stationary solutions, which they call gap waves, can be
regarded as the intermediate states between NBW and FGS. This
development is a great boost to such viewpoint.  However, doubt
always lingers as people know that a match between a NBW and the
related FGS can be quite bad for a different set of parameters, as
shown in Fig. \ref{First2}(a). Recently, we approached this
composition relation from a different angle \cite{Zhang}. There, we
presumed the existence of this composition relation and then
investigated how many conclusions could be drawn from it without any
computation. These conclusions were eventually verified through
numerical computation. In this way, we were able to make a firm
claim beyond doubt that the composition relation between NBWs and
FGSs exists. Note that a similar relation was pointed out by Carr
{\it et al.} \cite{Carr} for extensive periodic solutions and
solitons for a nonlinear system without periodic potential.

In this paper we explore in detail this composition relation between
NBWs and FGSs \cite{Zhang}. We not only offer more details on this
relationship but also generalize it to the systems with focusing
nonlinearity.  With this relation one can draw many conclusions
without any computation. For example, there are $n$ families of FGSs
in the $n$th band gap for the defocusing nonlinearity and the FGS of
the $n$th family has $n$ main peaks. All the conclusions will be
discussed in detail and be verified with extensive numerical
results.  Moreover, we have computed the Wannier functions from the
NBWs \cite{Liang} and compared them to the FGSs. We find that these
nonlinear Wannier functions match very well with the FGSs.  This
fact seems to suggest that the composition relation between NBWs and
FGSs is related to the exact relation between NBWs and nonlinear
Wannier functions. In addition, we have analyzed with different
methods the stabilities of the new-found FGSs  and the related gap
waves \cite{Alexander}. One method is linear stability analysis; the
other is the so-called nonlinear analysis by integrating the dynamic
equation with noise \cite{Yang}. Our numerical results show that not
all of these solutions are stable. The stability regions are marked
out.

There are numerical results in literature indicating that some FGSs
do not exist for all the values of propagation constant (or chemical
potential) in the linear band gap
\cite{Efremidis,Louis,Mayteevarunyoo}. Namely, there exists a kind
of cutoff. Yet it is not clear so far why there is such a cutoff.
 Here we show that
there indeed exists such a cutoff arising from the mixing of
different types of FGSs. This mixing can be intuitively viewed as a
result of a ``chemical reflection".

This paper is organized as follows. In Sec. II, we give a brief
description of our model equation and show how it is related to the
concrete systems. In Sec. III, we state the composition relation
between NBWs and FGSs and list all the predictions that can be made
with this relation. We then demonstrate that all the predictions are
valid.  In Sec. IV, the FGSs are compared to nonlinear Wannier
functions. They resemble each other very well, suggesting that the
composition relation is related to the well known relation between
Bloch waves and Wannier functions. In Sec. V, the composition
relation is applied to construct stationary solutions other than
Bloch waves with FGSs, such as gap waves and multiple periodic
solutions. In Sec. VI, the stabilities of FGSs and gap waves are
examined. In Sec. VII, we offer an explanation why FGSs do not exist
for all values of the propagation constant in the linear band gap.
Finally, we summarize our results in Sec. IX.

\section{Model equation}
We consider a one-dimensional nonlinear periodic system described by
\begin{equation}
      \label{NSE}
       i\frac{\partial\Psi}{\partial{z}}=-\frac{1}{2}\frac{\partial^2\Psi}{\partial{x^2}}+
       V(x)\Psi+\sigma|\Psi|^2\Psi,
\end{equation}
with $V(x)$ being a periodic function. Without loss of generality,
we will use $V(x)=\nu \cos(x)$ throughout this paper. The $\sigma$
in Eq. (\ref{NSE}) indicates the type of nonlinearity: $\sigma=1$
for the defocusing (or repulsive) case and $\sigma=-1$ for the
focusing (attractive) case.

In optics, Eq. (\ref{NSE}) describes light propagation along the $z$
direction in the presence of a periodic modulation in $x$ direction.
The periodic structure described by $V(x)$ can be experimentally
realized with waveguide arrays \cite{Christodoulides} or optical
inducing technology \cite{Fleischer}. As routinely used in
literature, $z$ and $x$ here are respectively scaled to diffractive
length and beam width.

In the context of the BEC system, Eq. (\ref{NSE}) gives the
description of a BEC in the one-dimensional optical lattice with $z$
being the time variable. In such case, Eq. (\ref{NSE}) has been
scaled as follows: $x$ is in units of $\Lambda/(2\pi)$, $z$ is in
units of $\hbar/(8E_{rec})$ and the strength of the optical lattice
$v$ is in units of $8E_{rec}$ with $m$ being the atomic mass,
$\Lambda$ the period of the lattice and
$E_{rec}=\hbar^{2}\pi^{2}/(2m\Lambda^{2})$ the recoil energy.

For stationary solutions in the form of
$\Psi(x,z)=\phi(x)\exp(-i\mu{z})$, Eq. (\ref{NSE}) is reduced to a
$z$-independent equation
 \begin{equation}
     \label{SNSE}
      -\frac{1}{2}\frac{\partial^2{\phi}}{\partial{x^2}}+\nu
      \cos(x) \phi+\sigma|\phi|^2\phi=\mu\phi.
  \end{equation}
Here, $\mu$ is referred to as propagation constant in optics.
Whereas in the BEC system, $\mu$ represents the chemical potential.
In general, NBW and FGS are the two basic types of stationary
solutions to Eq. (\ref{SNSE}). For the FGS which is localized in
space, we can define its norm $N$ as
   \begin{equation}\label{NormS}
      N=\int^{\infty}_{-\infty}|\phi(x)|^{2}dx.
   \end{equation}
By contrast, the NBW spreads over the whole space, hence it is only
meaningful to define its averaged norm $\mathcal {N}$ over one
period
   \begin{equation}\label{NormB}
      \mathcal {N}=\int^{2\pi}_{0}|\phi(x)|^{2}dx.
   \end{equation}
The so defined norm is proportional to the laser strength in optics,
or the number of atoms for a BEC.

To avoid confusion, hereafter we will present our results and
discussions in the framework of optics unless otherwise specified.
We first consider the defocusing case, $\sigma=1$.

\section{Composition relation in the defocusing case}

Without the nonlinear term, Eq. (\ref{SNSE}) is the well-known
Mathieu equation \cite{Jordan}. Its physical solutions are Bloch
waves defined by $\phi_{n,k}(x)=\exp(ikx)\psi_{n,k}(x)$ with
$\psi_{n,k}(x)=\psi_{n,k}(x+2\pi)$ \cite{Aschcoft}.  Here $k$ is the
Bloch wave vector and $n$ is the band index; the $\mu_{n}(k)$ form
Bloch bands as $k$ varies through the Brillouin zone (BZ). There
exist band gaps between different Bloch bands indexed by $n$, where
the physical solutions are forbidden.

With the addition of the nonlinear term, the physical solutions of
Eq. (\ref{SNSE}) become admissible in the linear band gaps. One such
typical solution is the gap soliton. Since the propagation constants
of gap soliton only take values inside the linear band gaps, no
linear counterpart exists for gap soliton. Among various gap
solitons, there is a particularly important class called FGSs, whose
main peaks locate inside one unit cell.

\begin{figure}[htb]
\includegraphics[bb=20 16 267 159, width=8.5cm]{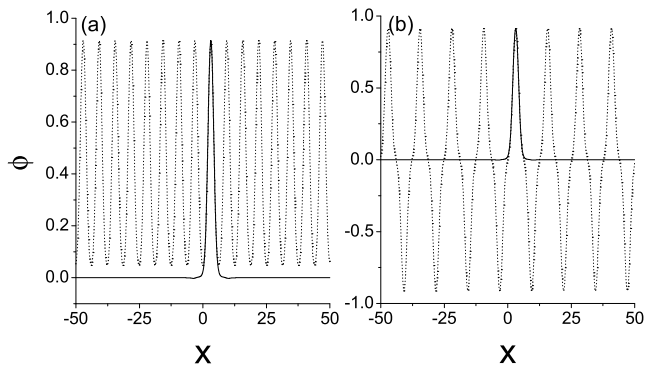}
\caption{NBWs (dotted lines) of the first nonlinear band and FGSs
(solid lines) in the first linear band gap for $\nu=1.5$ and
$\mu=-0.3$. NBW in (a) is at the center of BZ with $\mathcal
{N}=1.6908$; NBW in (b) is at the edge of the BZ with $\mathcal
{N}=1.6738$.} \label{First1}
\end{figure}
\begin{figure}[htb]
\includegraphics[bb=21 14 290 168,width=9cm]{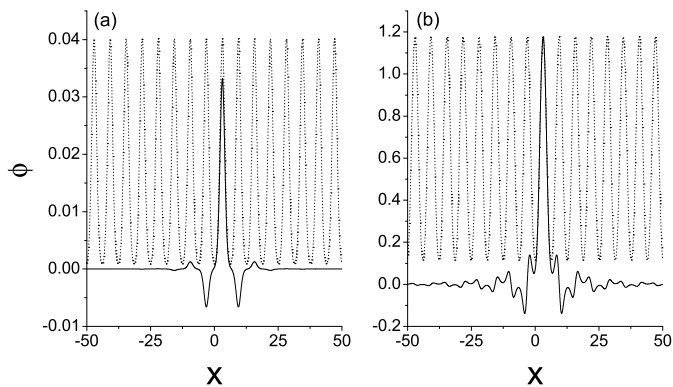}
\caption{NBWs (dotted lines) of the first nonlinear band and FGSs
(solid lines) in the first linear band gap for $\nu=1.5$. (a) NBW is
at the center of BZ with $\mathcal {N}=0.0027$ and $\mu=-0.92$ is
near the top of the first linear Bloch band; (b) NBW is at the edge
of the BZ with $\mathcal {N}=3.2194$ and $\mu=0.151$ is close to the
bottom of the second linear Bloch band.} \label{First2}
\end{figure}

Despite the nonlinearity, Eq. (\ref{SNSE}) still admits the Bloch
wave solutions. However, the nonlinear term will modify Bloch waves
and Bloch bands. For the defocusing case, the nonlinearity not only
changes the shapes of the Bloch bands but also moves them up into
the linear band gap \cite{Wu2}. Here, the strength of nonlinearity
is completely described by the norm $\mathcal {N}$. If $\mathcal N$
is lowered to zero, these nonlinear bands will move down and be
reduced to the linear Bloch bands. When $\mathcal N$ is increased,
bands will move up continuously without limit.

It is clear from the above discussions that there exist two types of
solutions to Eq. (\ref{SNSE}), NBW and FGS, for a given $\mu$ in the
linear band gap. In Fig. \ref{First1}, we have plotted both NBWs and
FGSs for $\mu=-0.3$ in the middle of the first linear band gap. Fig.
\ref{First1}(a) (or (b)) is for a NBW at the center (or edge) of the
BZ. In this figure, a nearly perfect match is found between the NBWs
and corresponding FGSs inside one unit cell.  These numerical
results therefore strongly suggest that FGS can be considered as the
building blocks for  NBWs at either the center or the edge of the
BZ. In other words, NBW at the center of the BZ can be viewed as an
infinite chain composed of FGSs while NBW at the edge of BZ is built
by FGSs with alternative signs.

However, such perfect match does not exist for all $\mu$ in the
linear band gaps. After checking various values of $\mu$, we find
that the match between the NBW and FGS is very good except in a
narrow region near the edge of the linear bands. Two typical results
are shown in Fig. \ref{First2}. Fig. \ref{First2}(a) is for the case
of $\mu$ near the top of the first Bloch band while Fig.
\ref{First2}(b) is for the case of $\mu$ close to the bottom of the
second Bloch band. It is evident from Fig. \ref{First2}(a) that for
a $\mu$ close to the edge of the first linear band, the NBW and the
FGS does not match well. This mismatch casts strong doubt on the
validity of the claim that a NBW can be regarded as an infinite
chain composed of FGSs.

In this work, the gap solitons are numerically obtained by using the
relaxation method in the coordinate space
\cite{Mayteevarunyoo,Kartashov} while the Bloch waves are
numerically found by applying the relaxation method in the Fourier
space \cite{Wu2}.

\subsection{Direct predictions from the composition relation}

We now take a different view at the above observed composition
relation between the FGSs and NBWs. We shall first presume the
existence of this composition relation and then try to draw as many
conclusions as possible. By verifying these conclusions, we will
justify this composition relation {\it a posteriori}. In this sprit,
the following predictions can be immediately drawn without any
computation.

(1) {\it For defocusing nonlinearity, there is no FGS in the
semi-infinite linear gap below the lowest Bloch band.}

As defocusing nonlinearity can be regarded as a result of repulsive
interaction, its addition to the system will increase the system
energy and therefore move the nonlinear Bloch bands up relative to
their linear counterparts. This means that there is no NBW for the
$\mu$ in the semi-infinite gap. According to the composition
relation, one can then conclude that there is no FGS in the
semi-infinite band gap.

(2) {\it There exist $n$ different families of FGSs in the $n$th
linear band gap for defocusing nonlinearity.}

\begin{figure}[tb]
\includegraphics[bb=22 19 256 128]{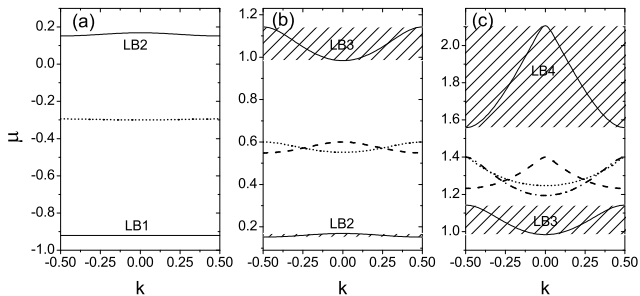}
\caption{Linear and nonlinear (defocusing case) Bloch bands for
$\nu=1.5$. Solid lines are linear bands, LB$i$ represents the $i$th
linear band. The widthes of linear bands are marked by shadow areas.
(a) The first nonlinear Bloch band (dotted line) in the first linear
band gap with $\mathcal {N}=1.6908$; (b) the first (dotted) and
second (dashed line) nonlinear bands in the second linear band gap
with $\mathcal {N}=4.8437$ and $1.5268$ respectively; (c) the first
(dotted), second (dashed), and third (dash-dotted line) nonlinear
bands in the third linear band gap with $\mathcal {N}=8.3478$,
$4.7363$, and $1.0242$ respectively.} \label{bandgap}
\end{figure}

In order to show this, we have plotted the linear and nonlinear
Bloch bands in Fig. \ref{bandgap}. Note that all the NBWs in the
same nonlinear Bloch band share the same nonlinearity $\mathcal
{N}$. As already discussed above, the nonlinearity $\mathcal {N}$
can move the nonlinear Bloch bands up. Therefore, the $m$th
nonlinear Bloch band can be lifted into the $n\geq m$ linear band
gaps. For example, as $\mathcal {N}$ increases, the first nonlinear
Bloch band can be lifted into the first, second, third, and all
other linear band gaps. This implies that there exists only one
nonlinear Bloch band ( the first nonlinear Bloch band) in the first
linear band gap [Fig.\ref{bandgap}(a)]; two nonlinear Bloch bands
(the first and second) in the second linear band gap
[Fig.\ref{bandgap}(b)]; three nonlinear Bloch bands (the first,
second, and third) in the third linear band gap
[Fig.\ref{bandgap}(c)]; and so on. In other words, there are $n$
different NBWs in the $n$th linear band gap. NBWs in the different
Bloch bands have different characters. With the composition
relation, one can immediately conclude that there are $n$ different
families of FGSs in the $n$th linear band gap.

In Ref.\cite{Wang2}, this rising nonlinear Bloch band by nonlinearity
was noted to be useful for analyzing gap waves (or truncated
Bloch waves).

(3) {\it In the $n$th linear band gap, the $m$th ($m<n$) family of
FGSs exists only above a threshold value of norm $N$ whereas the
$n$th family does not have such a value.}

Generally, one must increase nonlinearity $\mathcal {N}$ over a
critical value to move the $m$th nonlinear Bloch band up into the
$n$th ($n>m$) linear band gap while there is no such a critical
value to lift the $n$th nonlinear Bloch band into the $n$th linear
band gap. An example is shown in Fig. \ref{bandgap}(b). In order to
lift the first nonlinear band into the second linear band gap,
nonlinearity $\mathcal {N}$ must be beyond a threshold value while
there is no such a value to move the second nonlinear band into the
second linear band gap. This analysis, combined with the composition
relation, leads us to predict that there is a threshold value of
norm $N$ for the $m$th ($m<n$) family of FGSs in the $n$th linear
band gap while the $n$th family have no such a threshold value.

(4) {\it The $n$th family of FGSs has $n$ main peaks inside one unit
cell (or an individual well in the periodic potential).}

The linear Bloch waves in the $n$th linear Bloch band originate from
the $n$th bound state of an individual well of periodic potential.
Since the $n$th bound state has $n-1$ nodes, the linear Bloch waves
have $n$ main peaks in one unit cell. This character is shared by
the Bloch waves belonging to the $n$th nonlinear Bloch band.
Therefore, as the building blocks of NBWs, the $n$th family of FGSs
should have $n$ main peaks.

\begin{figure}[htb]
\includegraphics[bb=21 14 290 168,width=9cm]{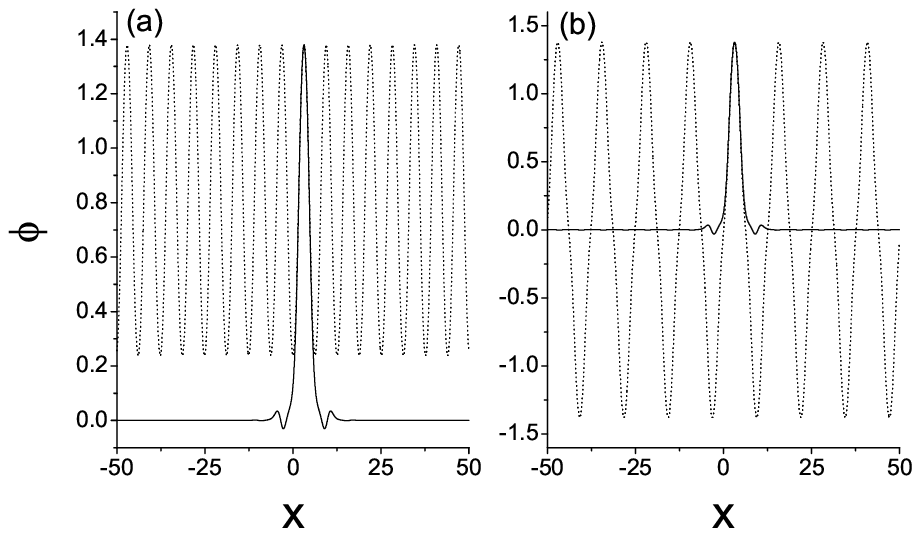}
\includegraphics[bb=21 14 290 168,width=9cm]{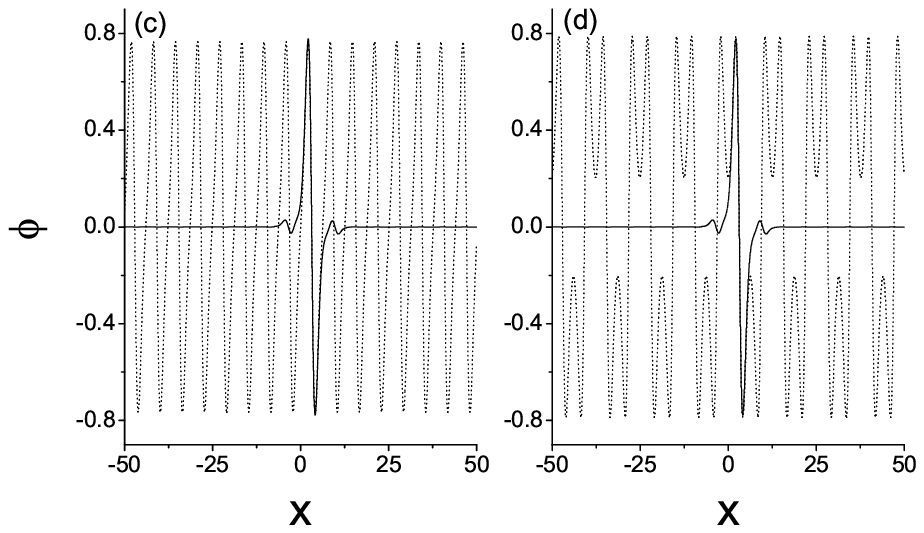}
\caption{NBWs (dotted lines) and FGSs (solid lines) in the second
linear band gap for $\nu=1.5$ and $\mu=0.6$. (a) A first-family FGS
and its corresponding NBW of the first Bloch band at the center of
the BZ with $\mathcal {N}=5.0598$; (b) a first-family FGS and its
corresponding NBW of the first Bloch band at the edge of the BZ with
$\mathcal {N}=4.8443$; (c)  a second-family FGS and its
corresponding NBW of the second Bloch band at the center of the BZ
with  $\mathcal {N}=1.5268$; (d) a second-family FGS and its
corresponding NBW of the second Bloch band at the edge of the BZ
with $\mathcal {N}=1.7417$.} \label{Second}
\end{figure}

\begin{figure}[htb]
\includegraphics[bb=19 14 290 168,width=9cm]{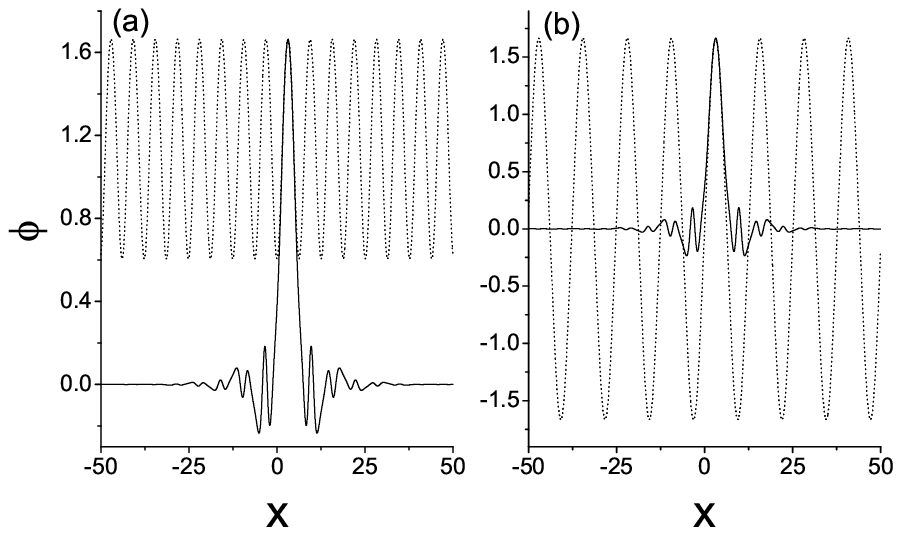}
\includegraphics[bb=19 14 290 168,width=9cm]{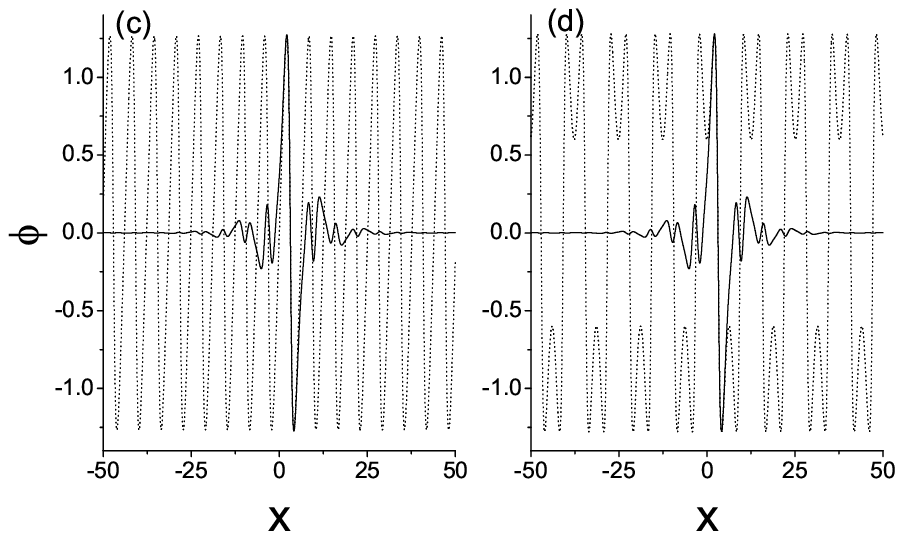}
\includegraphics[bb=19 14 290 168,width=9cm]{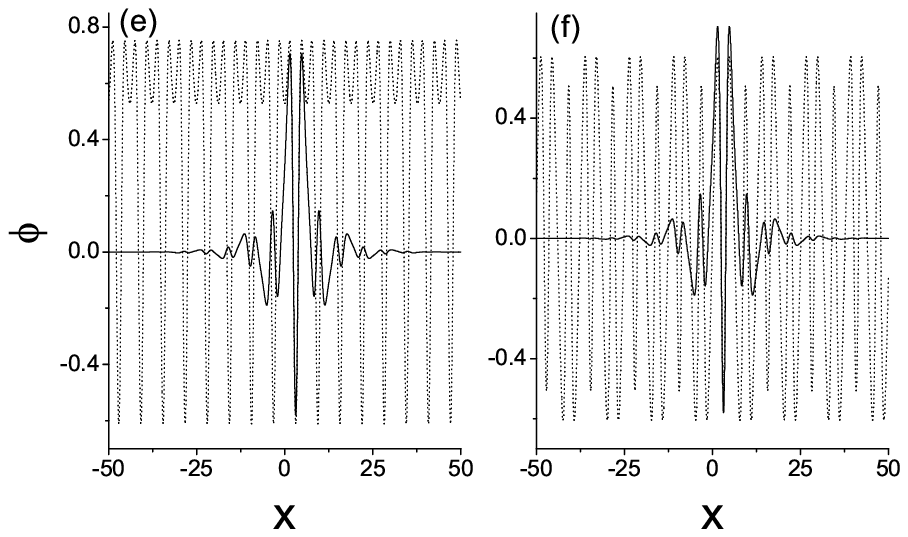}
\caption{NBWs (dotted lines) and FGSs (solid line) in the third
linear band gap for $\nu=1.5$ and $\mu=1.4$.  The solid lines are
for FGSs and dotted lines for NBWs.  (a,b) A first-family FGS; (c,d)
a second-family FGS; (e,f) a third-family FGS. (a,c,e) NBWs at the
center of the BZ in the first, second, and third bands with
$\mathcal {N}=9.2162, 4.7363$, and $2.052$, respectively; (b, d, f)
NBWs at the edge of the BZ in the first, second, and third bands
with $\mathcal {N}=8.3478, 5.6197$, and $1.0242$, respectively.}
\label{Third}
\end{figure}

\subsection{Verification of the predictions}
 In the following, we check the validity of the predictions listed
above.

(1) The first prediction is consistent with the well-known and
extensively proved fact that gap solitons do not exist in the
semi-infinite gap for defocusing nonlinearity
\cite{Efremidis,Louis}. As a result, this prediction can be
considered as the confirmation of a known result.

(2) We resort to the numerical computation to verify the second
prediction. As it is impossible to exam every linear band gaps, we
foucus on the second and third linear band gaps. We indeed find two
families of FGSs in the second linear band gap and three families of
FGSs in the third linear band gap. They are shown and compared to
the corresponding NBWs in Figs. \ref{Second} and \ref{Third}.

We note that the second family of FGSs are called subfundamental gap
solitons in literature \cite{Mayteevarunyoo}. This indicates that
people were not expecting other  FGSs to be found. In other words,
the existence of the third family of FGSs as shown in Figs.
\ref{Third} (e) and (f) is a surprise to many. Our results here also
show that all the families of FGSs should be regarded equally
fundamental as the corresponding NBWs are equally important.

\begin{figure}[!tb]
\includegraphics[bb=19 18 340 229,width=8cm]{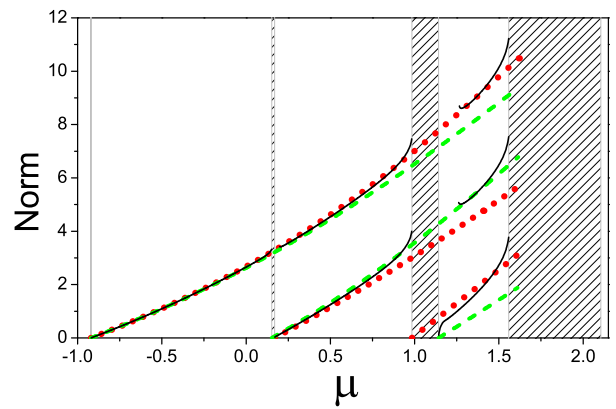}
\caption{(Color online) Norms of
 FGSs and NBWs as function of the $\mu$ for $\nu=1.5$. Shaded
areas are linear bands. Dotted(red), dashed(green), solid(black)
lines represent Bloch waves at the BZ center, Bloch waves at the BZ
edge and FGSs respectively.} \label{families}
\end{figure}

(3) In Fig. \ref{families}, we have plotted the $N$ as a function of
the $\mu$ for three different families of FGSs. It is clear from the
figure that in the second linear band gap the first family of FGSs
exists only when their $N>3.4140$ while the second family exists for
arbitrary small norm, having no threshold value. In the third linear
band gap, the first and second family exist only when their norms
$N>8.6105$ and $N>5.0152$, respectively. In contrast, the third
family of FGSs has no such threshold value. For comparison, we have
also plotted the norms $\mathcal N$ of the corresponding NBWs versus
$\mu$ in Fig. \ref{families}. These NBW norms $\mathcal N$ match
quite well with the FGS norms $N$.

\begin{figure}[!tb]
\includegraphics[bb=19 16 270 145,width=8cm]{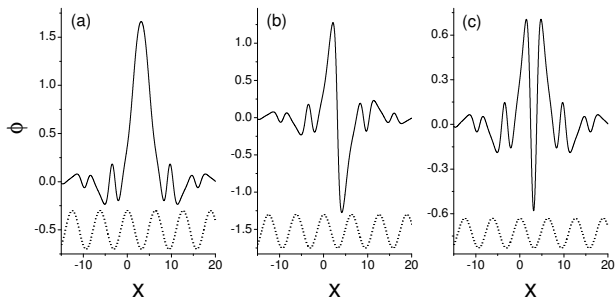}
\caption{Typical profiles of FGSs in the third linear band gap for
$\nu=1.5$, and $\mu=1.4$. (a) The first family; (b) the second
family; (c) the third family. Dotted lines in the bottom of each
figure mimic the periodic potential.} \label{Profiles}
\end{figure}

(4) We have also found that the number of main peaks of a FGS in a
well is just what we have expected. In order to demonstrate this
clearly, we have plotted Fig. \ref{Profiles} the three families of
FGSs in the third linear band gap along with the periodic potential.
It is clear that the first family of FGSs has one main peak in an
individual well [Fig. \ref{Profiles}(a)]; the second and third
families of FGSs have two and three main peaks in a unit cell,
respectively, as shown in Figs. \ref{Profiles}(b) and (c).

\section{Nonlinear Wannier Function and Fundamental gap soliton}

\begin{figure}[!tb]
\includegraphics[bb=20 17 145 229,width=7cm]{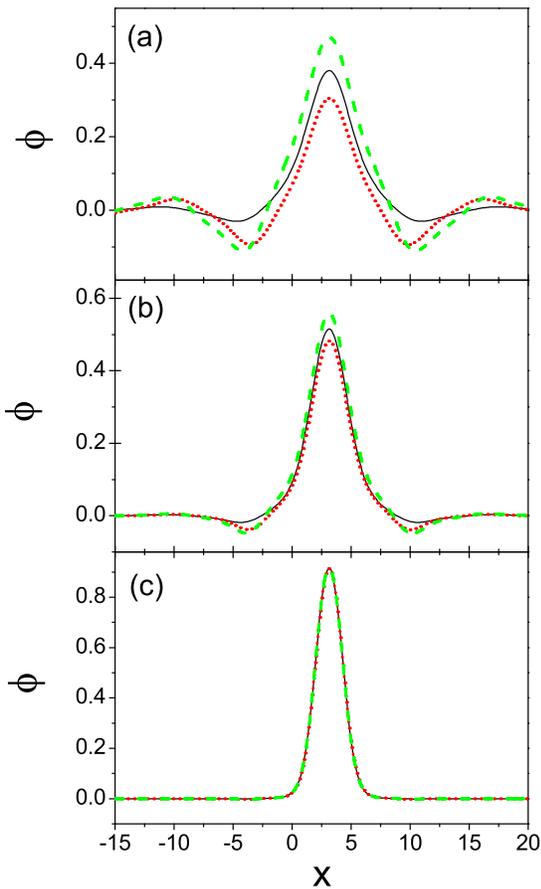}
\caption{(Color online) Nonlinear Wannier functions for the first
nonlinear Bloch band and the corresponding FGSs. Solid (black) lines
are for Wannier functions, dashed (green) lines are for the FGSs
corresponding to the NBWs at the  band edge, and dotted (red) lines
are for the FGSs corresponding to the NBWS at the band center. (a)
$\nu=0.2$ and $\mathcal {N}=0.5027$; (b) $\nu=0.4$ and $\mathcal
{N}=0.7540$; (c) $\nu=1.5$ and $\mathcal {N}=1.6908$.}
\label{Wannier}
\end{figure}
For a linear periodic system, there is another well-known function,
Wannier function, which is localized in space. In one dimension, the
Wannier function $W_{n}(x)$ is related to the Bloch wave
$\phi_{n,k}$ as follows \cite{Aschcoft}
\begin{equation}\label{Wannier}
\phi_{n,k}(x)=\sum_j e^{ikx_j}W_{n}(x-x_j)\,,
\end{equation}
where $x_j$ is the location of the $j$th well in  the periodic
potential and the summation runs over all  potential wells. The
relation of Eq. (\ref{Wannier}) is still valid for nonlinear
periodic systems and the corresponding Wannier function can be
called nonlinear Wannier function \cite{Liang}. At either the center
or the edge of the BZ, Eq. (\ref{Wannier}) becomes
\begin{equation}
\phi_{n,\pm}(x)=\sum_j (\pm)^j W_{n}(x-x_j)\,,
\end{equation}
where $+$ is for the center and $-$ is for the edge.  This relation
is very similar to the composition relation between NBWs and FGSs
that we have just established. Hence, a question naturally arises
whether the FGSs bear any relation to the Wannier functions.

In Fig. \ref{Wannier}, we have plotted nonlinear Wannier functions
and the corresponding FGSs for the first Bloch band together for
comparison. The Wannier functions are computed from the Bloch waves
in a standard way \cite{Liang}. For the nonlinearities $\mathcal N$
considered here, the first nonlinear Bloch band lies completely in
the first linear band gap. As a result, there are two different FGSs
for this band, one corresponding to the NBW at the BZ center and the
other to the NBW at the BZ edge. Both of the FGSs are plotted in
Fig. \ref{Wannier} and are found to match the Wannier functions very
well. In fact, the match gets better as the periodic potential gets
stronger. Since the Wannier function is normalized, we have scaled
them by a factor $\sqrt{N}$ for comparison with the FGSs in Fig.
\ref{Wannier}.

The excellent match between the FGSs and the Wannier functions
suggests that they are related. This relationship may be intuitively
understood in the following way. Although a Bloch wave is a solution
of a periodic system, the Wannier function is not. In a linear
periodic system, any localized wave function, including a Wannier
function, will spread in space. In a nonlinear periodic system, it
seems that the Wannier function can be modified slightly and become
a solution of the system in the form of a FGS.

\section{Fundamental gap solitons: building blocks for
stationary solutions}

The composition relation between NBWs and FGSs suggests that the
FGSs are really fundamental and can be viewed as building blocks for
other stationary solutions to a nonlinear periodic system, such as
high-order gap solitons. Our numerical results and existing results
in literature \cite{Alexander,Carr,Kartashov,Smirnov} fully support
this view. In the following, we show a few examples.

\subsection{Gap waves}

\begin{figure}[htb]
\includegraphics[bb=12 11 178 228,width=8.5cm]{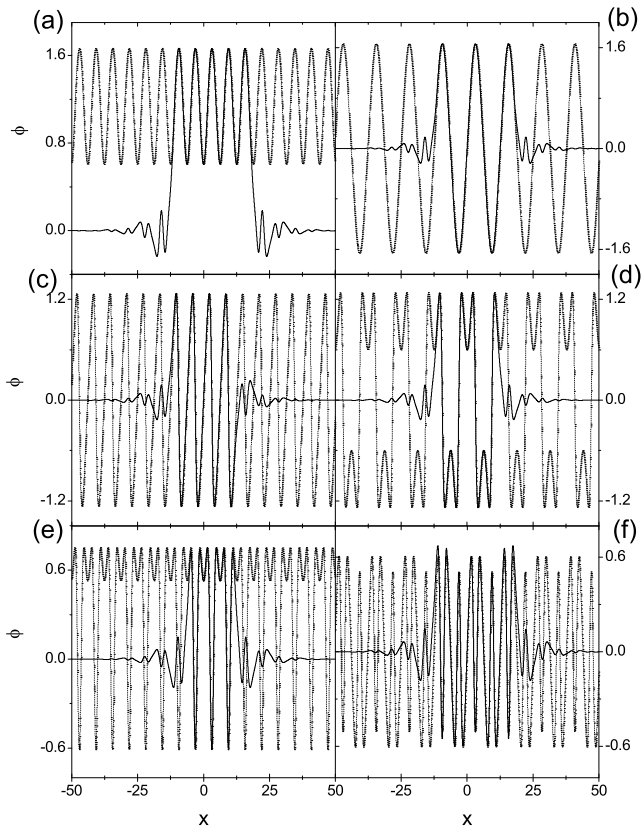}
\caption{Gap waves (solid lines) and the corresponding NBWs (dotted
lines) in the third linear gap for $\nu=1.5$ and $\mu=1.4$. The NBWs
in (a), (c), and (e) are at the BZ center with $\mathcal {N}=9.2162,
4.7363$, and $2.052$ respectively; the NBWs at the BZ edge in (b),
(d), and (f) have $\mathcal {N}=8.3478, 5.6197$, and $1.0242$
respectively.} \label{GW1}
\end{figure}
In the first example, we construct new solutions by putting finite
number of FGSs together. Such solutions are called high order gap
solitons in Refs. \cite{Kartashov,Smirnov}. There are numerous ways
to build these high order gap solitons. For instance, one can use
FGSs from different families and put them together with either the
same phase or the opposite phase. Here we mainly focus on a class of
high order gap solitons called gap waves by Alexander {\it et al.}
since they can be viewed as truncated NBWs \cite{Alexander}. For gap
waves, all the constituent FGSs come from the same family. These gap
waves can be viewed as intermediate states between NBWs and FGSs.
With one FGSs, two different classes of gap waves can be
constructed. The first class of gap waves, which we call GW-I, are
built by putting FGSs side by side similar to the NBW at the center
of the BZ. The second class, called GW-II, are composed of FGSs
pieced together with alternative signs similar to the NBW at the BZ
edge. Some typical gap waves in the third linear band gap and the
corresponding NBWs are plotted in Fig. \ref{GW1}. Figs.
\ref{GW1}(a), (c), and (e) are GW-I composed of  the first, second,
and third families of FGSs respectively; Figs. \ref{GW1}(b), (d),
and (f) are GW-II. Note that gap waves and the corresponding NBWs
have the same propagation constant $\mu$.

\begin{figure}[!tb]
\includegraphics[bb=16 15 246 177,width=8.5cm]{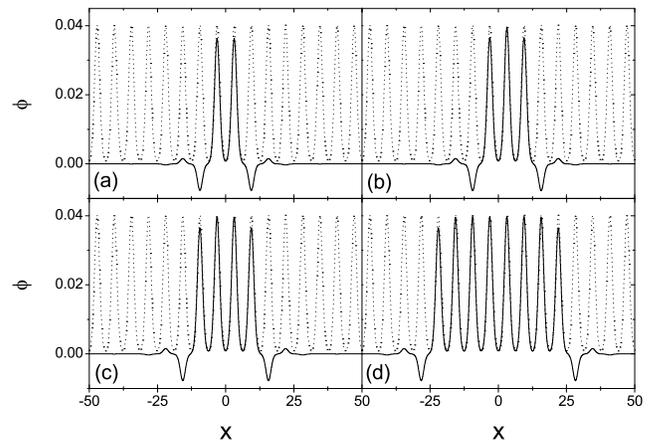}
\caption{Gap waves composed of different number of FGSs for
$\nu=1.5$ and $\mu=-0.92$. Solid  lines are for gap waves; dotted
line is for the NBW with $\mathcal {N}=0.0027$. } \label{GW2}
\end{figure}

We have also plotted gap waves composed of different number of FGSs
in Fig. \ref{GW2} along with the corresponding NBW. The parameters
in this figure are the same with ones in Fig. \ref{First2}(a), where
the mismatch between the FGS and the NBW is obvious. Fig. \ref{GW2}
shows an interesting trend that the match between the gap waves and
the NBW improves as the number of FGSs increases. This can be viewed
as another supporting evidence for the composition relation.

\subsection{Multiple periodic solutions}

\begin{figure}[!tb]
\includegraphics[bb=18 21 206 222,width=8.5cm]{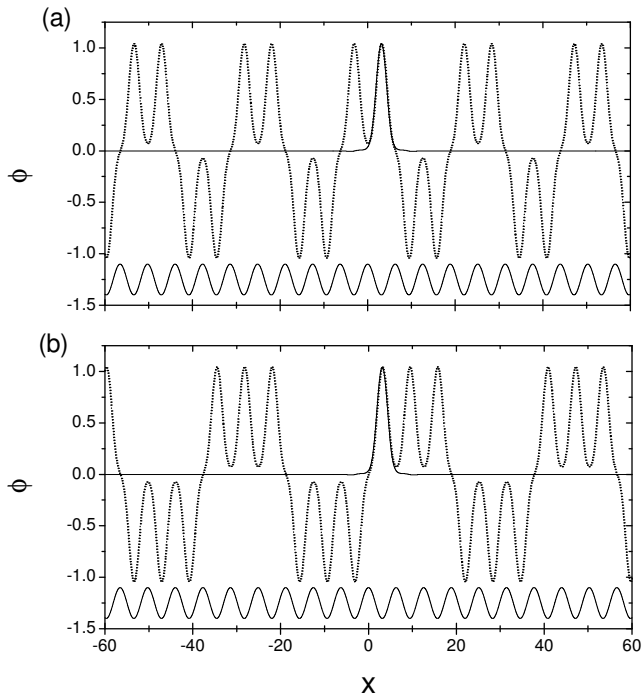}
\caption{Multiple periodic solutions for $\nu=1.5$, $\mu=-0.1$. (a)
period doubled solution; (b) triple periodic solution. Dotted lines
are multiple periodic solutions, solid lines are corresponding FGS.
Lines in the bottom of each figure mimic periodic potential.}
\label{Multiple}
\end{figure}

Multiple periodic solutions are extensive states like Bloch waves,
but with multiple periods. Mathematically, they are defined as
$\phi(x)=\exp(ikx)\psi_{k}(x)$ with $\psi_{k}(x)=\psi_{k}(x+2p\pi)$
and $p>1$ being an integer \cite{Machholm}.

The composition relation between FGSs and NBWs can be generalized to
construct multiple periodic solutions. Here we show two typical
kinds of multiple periodic solutions in Fig. \ref{Multiple}: one is
double periodic solution [Fig. \ref{Multiple}(a)]; the other is
triple periodic solution [Fig. \ref{Multiple}(b)]. The double
periodic solution is constructed with the pattern ``++- -" and the
triple periodic solution is built with the pattern ``+++- - -". One
can build other multiple periodic solutions with other patterns. In
fact, one can also build the multiple periodic solutions with FGSs
from different families. The odd-periodic solutions were speculated
to exist in Ref. \cite{Machholm}.

\subsection{Composition relation in the presence of a loop}

\begin{figure}[htb]
\includegraphics[bb=21 16 178 216,width=8cm]{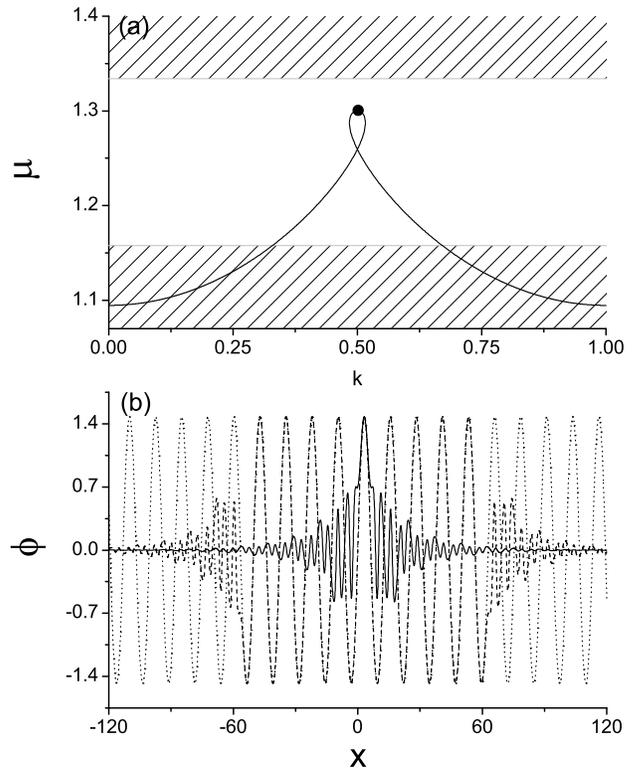}
\caption{(a) Loop structure of the first nonlinear band lying in the
third linear band gap for $\nu=1$ and $\mathcal {N}/2\pi=1.1344$.
Shadow areas are the linear bands and the blank area is the third
band gap; (b) NBW (dotted line) for the solid point in (a) with
$\mu=1.3$ and the corresponding gap wave (dashed line) and FGS
(solid line) in the third band gap.}\label{loop}
\end{figure}

We have seen in Fig. \ref{bandgap} that all the nonlinear bands move
up with the increasing defocusing nonlinearity. It is known that a
more dramatic change can happen when the nonlinearity is large
enough: loop structures emerge at the BZ edge for the first band and
at the BZ center for the second band \cite{Wu2,Diakonov}. The
critical value of the nonlinearity for the loop to appear is
${\mathcal N}>2\pi\nu$ \cite{Wu2}. Our studies show that the
composition relation still holds in the presence of such a loop. As
an example, we demonstrate this relation for a NBW sitting at the
loop top of the first band in Fig. \ref{loop}. In this figure, the
looped nonlinear Bloch band is already in the third linear band gap.

\section{stability properties of Fundamental gap solitons and gap waves}
\begin{figure}[b]
\includegraphics[bb=17 16 226 137,width=8cm]{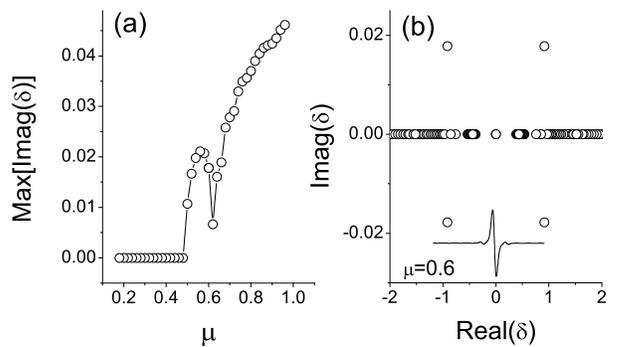}
\caption{Stability  of the second family FGS in the second linear gap.
(a) Maximum imaginary part of $\delta$ as a function of $\mu$; (b)
the eigenvalue plane of $\delta$ at $\mu=0.6$. The inset is
the profile of the corresponding gap soliton. $\nu=1.5$.} \label{S1}
\end{figure}
A question that must be asked for a stationary solution (or a fixed
point solution) in a nonlinear system is whether the solution is
stable. The unstable solution is very sensitive to small
perturbations. The stabilities of the NBWs have been discussed
thoroughly in Ref.\cite{Wu1,Dong}. In the following, we shall focus
only on FGSs and gap waves.

We use two different approaches to examine the stabilities of the stationary
solutions. The first is called linear stability analysis. It is done by adding a perturbation term to a
known solution
   \begin{equation}
     \Psi(z,x)=\Big[\phi(x)+\Delta\phi(z,x)\Big]\exp(-i\mu z),
     \label{perturb}
   \end{equation}
where $\Delta\phi(z,x)=u(x)\exp(i\delta z)+w^{*}(x)\exp(-i\delta^{*}
z)$ is the perturbation and $\phi(x)$ is the stationary solution.
Plugging Eq. (\ref{perturb}) into Eq. (\ref{NSE}) and keeping only
the linear terms, we obtain as follows
\begin{equation}
\begin{pmatrix}
\mathcal {L}&-\phi^2\\
{\phi^{*}}^{2}&-\mathcal {L}
\end{pmatrix}
\begin{pmatrix}
u\\
w
\end{pmatrix}
=\delta
\begin{pmatrix}
u\\
w
\end{pmatrix},
\label{LinearStability}
\end{equation}
with $\mathcal {L}=\frac{1}{2}\frac{d^2}{dx^2}-\nu
\cos(x)-2|\phi|^2+\mu$. In Eq. (\ref{LinearStability}), if the
eigenvalue $\delta$ has imaginary parts, the solution of $\phi(x)$
is unstable; otherwise, the solution is stable.

In the second method, the perturbed solution in Eq. (\ref{perturb})
is used as the initial condition for Eq. (\ref{NSE}). Its evolution
is then monitored numerically. If its deviation from $\phi(x)$ grows
as the system evolves, the solution $\phi(x)$ is unstable; it is
stable otherwise. The stability checked by this method is called
nonlinear stability \cite{Yang}.

\subsection{Fundamental gap solitons}

Our linear stability analysis shows that the first family of FGSs in
the first and second band gaps are stable consistent with Ref.
\cite{Louis}. However, they become unstable in a small area near the
band edges in the third gap.

\begin{figure}[htb]
\includegraphics[bb=18 18 223 130,width=8cm]{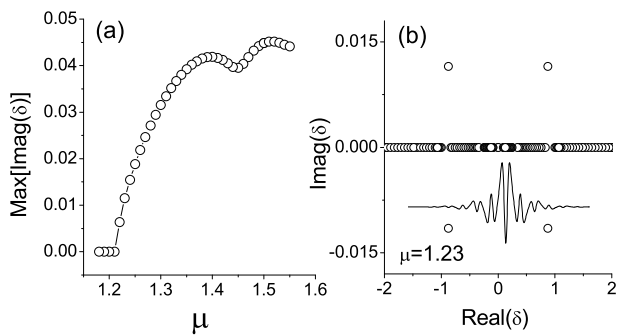}
\caption{Stability of the third family of FGSs in the
third gap. (a) Maximum imaginary part of $\delta$; (b) the
eigenvalue plane of $\delta$ at $\mu=1.23$.  The inset is the
corresponding gap soliton.  $\nu=1.5$.}  \label{S2}
\end{figure}

The stability of the second family of FGSs in the second linear band
gap is shown in Fig. \ref{S1}(a), where the stability is measured by
the maximum value of the imaginary part of the eigenvalues $\delta$.
If the maximum value is zero, the solution is stable; otherwise, it
is unstable. It is clear from Fig. \ref{S1}(a) that this family of
FGSs is stable when their $\mu$ are smaller than a critical value
near 0.5. For other values of $\mu$ above this critical value, the
solitons become unstable. Fig. \ref{S1}(b) are the eigenvalues for a
soliton illustrated in the inset. We see that the eigenvalues
$\delta$ are mostly real and become complex only in a very small
region.  The stability of the third family in the third linear band
gap is shown in Fig. \ref{S2}. The result is very similar to that of
Fig. \ref{S1} except that the stable region is much smaller. Fig.
\ref{S2}(b) shows an example of the eigenvalue plane.

\begin{figure}[htb]
\includegraphics[width=7cm,height=9cm,angle=270]{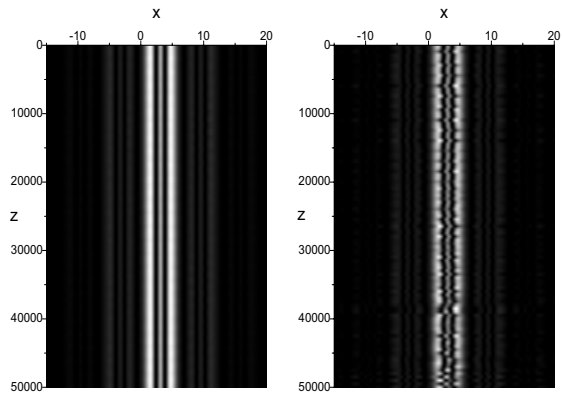}
\caption{Propagations of gap solitons in the presence of Gaussian
distributed random noise with variance $\sigma^{2}=0.01$. (a)
Evolution of a third family FGS at $\mu=1.21$ which is  stable as
indicated by the linear analysis results in Fig.\ref{S2}(a); (b)
evolution of an unstable third family FGS at $\mu=1.23$, which is
unstable as indicated by the linear analysis results in
Fig.\ref{S2}(a) (Max[Imag($\delta$)]=0.0115).} \label{S3}
\end{figure}

To double check the stability results, we have propagated perturbed
FGSs by numerically solving Eq. (\ref{NSE}) with the split-step
Fourier method. Gaussian distributed random noise is added to FGSs
for the initial wave function. The propagation results shown in Fig.
\ref{S3} agree with our linear stability analysis. Fig. \ref{S3}(a)
is the propagation of a stable solution while Fig. \ref{S3}(b) is
for an unstable solution.

The above results show that the newly-found third family FGS can be
stable and therefore should be observable in experiment. Usually in
order to observed gap solitons experimentally, initial input beam
profile should be close to the desired soliton profiles
\cite{Mandelik,Neshev}. We propose to observe the third family FGS
using two localized laser beams, whose wavelength is much shorter
than the period of a waveguide, to form an interference pattern with
three large peaks in a unit cell of the periodic waveguide.

\subsection{Gap waves}
\begin{figure}[htb]
\includegraphics[bb=21 19 208 134,width=8cm]{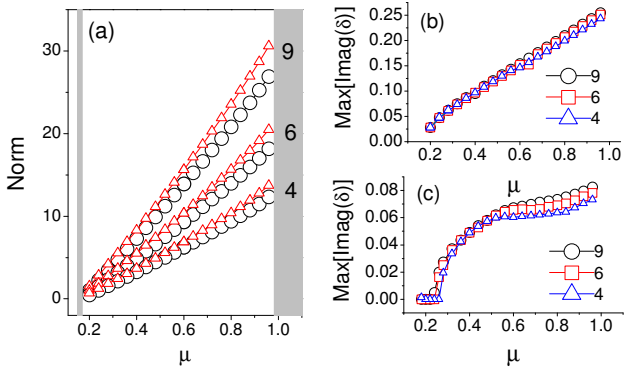}
\caption{(Color online) Stabilities of  gap waves composed
of the second family of FGSs in the second gap. (a) the families of
gap waves. Circles are for GW-I while triangles for GW-II. Shadow
areas are the linear bands; (b) maximum imaginary part of $\delta$
for GW-I; (c) maximum imaginary part of $\delta$ for
GW-II.}\label{S4}
\end{figure}

\begin{figure}[htb]
\includegraphics[width=6cm,height=8cm,angle=270]{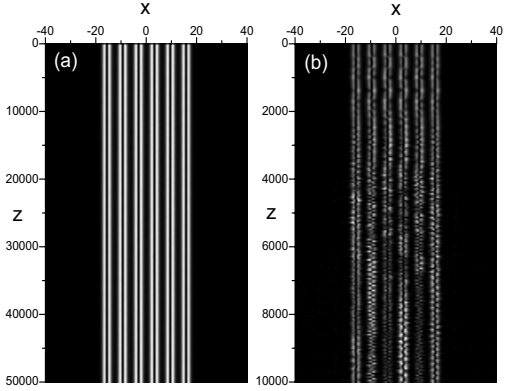}
\caption{Propagations of gap waves in the presence of a Gaussian
distributed random noise with variance $\sigma^{2}=0.01$. (a)
propagation of GW-II containing six FGSs at $\mu=0.24$, which is
stable indicated by the linear analysis result shown in Fig.\ref{S4}(c); (b)
propagation of GW-II with six second family FGSs at $\mu=0.26$,
which is unstable as shown in Fig.\ref{S4} (c)
(Max[Imag($\delta$)]=0.0162). Note the different scale of $z$ in (a)
and (b).}\label{S5}
\end{figure}

The stability of gap waves is also analyzed. In general, the GW-I's
corresponding to the first band are stable in the first and second
gaps but unstable in the third gap. The stability of gap waves
composed of the second family of FGSs in the second gap is shown in
Fig. \ref{S4}. These gap waves contain either 4, 6, or 9 FGSs. These
gap waves are characterized by the norms in Fig. \ref{S4}(a). Fig.
\ref{S4}(b) demonstrates that GW-I are always unstable. GW-II are
stable in a small regime near the top of the second band, but they
are unstable for other values of $\mu$ as shown in Fig. \ref{S4}(c).
The propagation of the gap waves with noise confirms our stability
analysis as shown in Fig. \ref{S5}. As the three curves fall almost
on top of each other in Figs. \ref{S4}(b,c), we find that the
stability of gap waves are independent of how many FGSs they have.
Our analysis shows that gap waves other than the types discussed
above are unstable.

\section{composition relation in the focusing case}
We have concentrated on the defocusing case. We now turn to the
focusing case. For focusing nonlinearity,  Bloch bands and NBWs
still exist. Unlike the defocusing case, focusing nonlinearity
causes nonlinear bands move down.  As a result, the predictions made
from  the composition relation are different from the defocusing
case. (1) There exist infinite number of families of FGSs in the
semi-infinite and finite  linear band gaps. It is because infinite
number of bands can move into a given linear band gap with
increasing focusing nonlinearity. (2) In the $n$th linear gap
($n\geq 0$ with $n=0$ for the semi-infinite gap), the $n$th family
and other lower order fmailies of FGSs do not exist. (3) In the
$n$th linear gap, only the ($n+1$)th family FGSs exist for an
arbitrary small values of norm while all other families of FGSs
exist only for norms above certain threshold values.

\begin{figure}[!tb]
\includegraphics[bb=18 16 215 168,width=8cm]{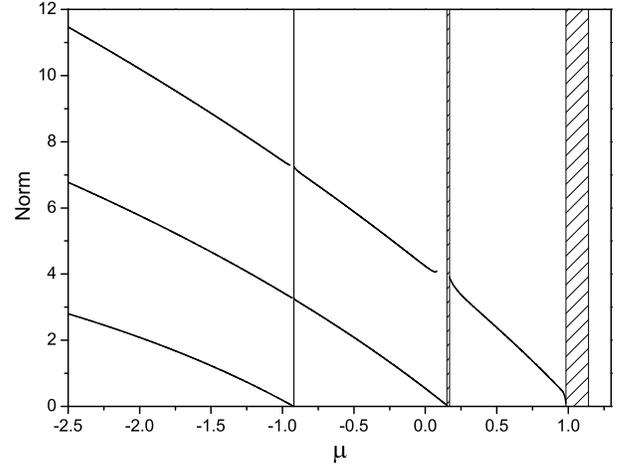}
\caption{Norms of fundamental gap solitons as functions of $\mu$ for
focusing nonlinearity.   Shaded areas are for linear bands; lank
areas are for gaps. $\nu=1.5$.} \label{familiesAttract}
\end{figure}

These predictions are confirmed by our numerical computation for the
first three bands and the corresponding three band gaps. The results
are summarized in  Fig.\ref{familiesAttract}, where the norms of
different FGSs are plotted as functions of $\mu$. As shown in this
figure, corresponding to these three bands,  there are three
families of FGSs in the semi-infinite linear band gap,  two families
of FGSs in the first linear band gap, and one family of FGSs in the
second band gap.  In other words, there exist no first family of
FGSs in the first linear gap and there exist no first and second
families of FGSs in the second gap. Another feature in the figure is
the threshold values of norm for some families of FGSs. In the
semi-infinite gap, the threshold value for the second family of FGSs
is $3.2824$ and for the third family $7.2940$. In contrast, the
first family have no such threshold value. In the first gap, the
threshold value of norm for the third family of FGSs is $4.0616$
while the second family has no threshold value. The match between
the norms of NBWs and FGSs is similar to that in the defocusing
case.

Combining with the results for the defocusing case, we have an
interesting observation: in the $n$th linear Bloch band gap, the first
$n$ families of FGSs exist for the defocusing nonlinearity and the other
families ($(n+1)$th, $(n+2)$th, $\cdots$) exist for the focusing nonlinearity.

\section{Chemical reflection of FGS}

One may have noticed in Figs. \ref{families} and
\ref{familiesAttract} that some families of FGSs do not exist for
all values of the $\mu$ in the linear band gaps. For example, in
Fig. \ref{families}, the second and third families of FGSs in the
third linear band gap do not exist for $\mu$ near the edge of third
linear Bloch band. The $N$-$\mu$ curves for these two families end
at $\mu=1.2658$ and $\mu=1.2612$, respectively, which are away from
the right edge of third linear band at $\mu=1.1422$. This cut-off
phenomenon was noticed before \cite{Alexander}. However, to our best
knowledge, no one is sure why this cut-off happens. In the
following, we show that this cut-off is caused by the mixing of
different types of FGSs, which can be intuitively viewed as a result
of a ``chemical reflection". It will be discussed for both
defocusing and focusing nonlinearities.

\begin{figure}[t]
\includegraphics[bb=17 18 244 203,width=8cm,height=8cm]{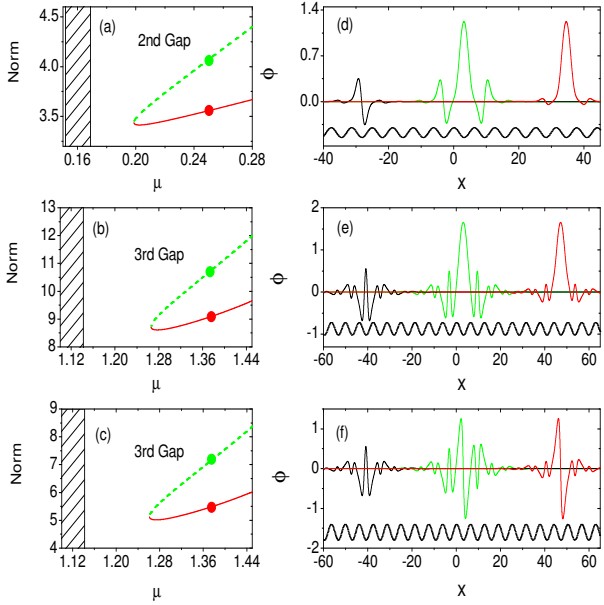}
\caption{(Color online) Chemical reflection of FGSs into high order
gap solitons in the ($N, \mu$) plane for defocusing case with
$\nu=1.5$. (a) The $N$-$\mu$ curves for the first family of FGSs in
the second gap (solid-red line) and the higher order solitons
(dashed-green line) composed of one first-family FGS and two
second-family FGSs; (b) the $N$-$\mu$ curves for the first family of
FGSs in the third gap (solid-red line) and the solitons
(dashed-green line) consisting of one first-family FGS and two
third-family FGSs; (c) the $N$-$\mu$ curves for the second family of
FGSs (solid-red line) and the solitons (dashed-green line) composed
of one second-family FGS and two third-family FGSs. (d, e, f)
soliton profiles corresponding to the labeled points in (a, b, c),
respectively. Green lines are for profiles of high order solitons,
red lines are FGSs, and black  lines are profiles of the second
family FGS in (d) and the third family FGS in (e) and (f). The lines
in the bottom represent periodic potential. } \label{Resonance}
\end{figure}

We consider first the defocusing case. We have re-plotted the
$N-\mu$ curves (solid lines) in Figs. \ref{Resonance}(a,b,c), where
the cut-offs exist. In these three figures, we have also plotted the
$N$-$\mu$ curves (dashed lines) for three different classes of
high order gap solitons. Interestingly, they are connected smoothly to
the curves for the FGSs. As illustrated in Figs.\ref{Resonance}(d,e,f),
we find after careful examination  that the
high order gap solitons for the dashed curve in
Fig.\ref{Resonance}(a) are composed of a first-family FGS sitting in
one site and two second-family FGSs sitting in two neighboring
sites,   the high order solitons in Fig.\ref{Resonance}(b) composed
of a first-family FGS  and two third-family FGSs, and the high order
solitons in Fig.\ref{Resonance}(c) consists of a second-family FGS
and two third-family FGSs.

To help us understand the turning $N-\mu$ curves, we have developed
an intuitive picture to visualize this result. We use the case
in Fig. \ref{Resonance}(a) as an example.
If one imagines an ``atom" moving along the lower curve for the FGSs in
the ($N,\mu$) plane in Fig. \ref{Resonance}(a), this ``atom" gets
reflected by the ``repulsive walls" of the second linear band.
Moreover, a ``chemical reaction" occurs during the collision between
this ``atom" of FGS and the ``wall", which may be viewed as
a crystal   made of ``atoms" of the second-family FGS. The result of
this reaction is
that the ``atom" changes its nature to a ``molecule" of high order
soliton by picking up two second-family FGSs from  the
second linear band. This ``chemical reaction" similarly occurs in
Fig. \ref{Resonance}(b,c). Based on this intuitive picture, we
call this cut-off phenomenon in Figs. \ref{Resonance}(a,b,c)
chemical reflection.

Note that a similar turning $N-\mu$ curve  was also found for gap
vortexes in Ref.\cite{Wang} and gap waves in Ref.\cite{Wang2}.

\begin{figure}[!tb]
\includegraphics[bb=17 18 266 202,width=8cm]{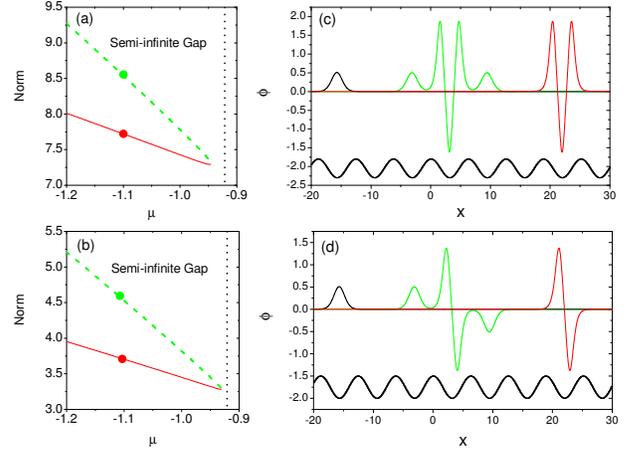}
\caption{Chemical reflection of the second and third families of
FGSs with higher order gap soliton in the semi-infinite gap for
focusing case. $\nu=1.5$. (a) the third family of FGSs (solid-red
line) and higher order gap solitons (dashed-green line) composted of
one third-family FGS and two first-family FGSs. (b) the second
family of FGSs (solid-red line) and higher order gap solitons
(dashed-green line) composted by one second-family FGS and two
first-family FGSs. Dotted lines in (a) and (b) are the first linear
band. (c) and (d) soliton profiles corresponding to labeled point in
(a) and (b) respectively. Green lines are for profiles of high order
solitons, red  lines are FGSs, and black lines are profiles of the
first family FGS. The lines in the bottom represent periodic
potential. }\label{CutoffAttrac}
\end{figure}

The cut-off phenomenon for the focusing case as shown in Fig.
\ref{familiesAttract} can be similarly be viewed as the result of
the chemical reflection. In the focusing case, as shown in Fig.
\ref{familiesAttract}, except the lowest family of FGSs in each
linear gap, all other families have cut-offs in the  propagation
constant $\mu$. The cut-off phenomenon can be similarly viewed as
the result of the chemical reflection as demonstrated  in Fig.
\ref{CutoffAttrac}. Fig. \ref{CutoffAttrac}(a) is for the third
family of FGSs in the semi-infinite gap, whose $N$-$\mu$ curve is
found to be connected smoothly to a class of high order gap solitons
composed of a third family FGS in one site with two FGSs of the
first family in its neighboring sites [see Fig.
\ref{CutoffAttrac}(c)]. The case for the second family of FGSs in
the semi-infinite gap is shown in Fig. \ref{CutoffAttrac}(b), where
the high order solitons consist of  one FGS of the second family and
two FGSs of the first family [see Fig. \ref{CutoffAttrac}(d)].

\section{conclusion}

In conclusion, we have demonstrated that a composition relation
exists between FGSs and NBWs for both defocusing and focusing
nonlinearities. Based on the composition relation, we have drawn
many conclusions about the properties of FGSs directly from Bloch
band gap structures without any computation. All the predictions
have been examined and confirmed. All our studies point to one
important conclusion that the FGSs are really fundamental and they
serve as building blocks for other stationary solutions in
one-dimensional nonlinear periodic systems.

\section{Acknowledgments}
We thank Zhiyong Xu for the helpful discussions. This work was
supported by the NSF of China (10825417) and the MOST of
China (2005CB724500,2006CB921400). L.Z.X. is supported by
the IMR SYNL-T.S. K$\hat{e}$ Research Fellowship.


\begin{thebibliography}{99}

\bibitem{Lederer} {F. Lederer, G. I. Stegemanb, D. N. Christodoulides,
G. Assanto, M. Segev, and Y. Silberberg, Phys. Rep. \textbf{463}, 1
(2008).}

      \bibitem{Morsch} O. Morsch and M. Oberthaler, Rev. Mod. Phys. \textbf{78},
     179 (2006).

     \bibitem{Christodoulides} D. N. Christodoulides, F. Lederer, and Y.
      Silberberg, Nature (London) \textbf{424}, 817 (2003).

      \bibitem{Eisenberg} H. S. Eisenberg, Y. Silberberg,  R. Morandotti, A. R. Boyd, and J.
      S. Aitchison, Phys. Rev. Lett. \textbf{81}, 3383 (1998).

      \bibitem{Fleischer} J. W. Fleischer, M. Segev, N. K. Efremidis, and
      D. N. Christodoulides, Nature (London) \textbf{422},147 (2003).

     \bibitem{Brazhnyi}V. A. Brazhnyi and V. V. Konotop, Mod. Phys. Lett. B
     \textbf{18}, 627 (2004).

    \bibitem{Aschcoft}Neil W. Aschcoft and N. David Mermin, Solid state
    phsyics (Saunders College publishing, 1976).

     \bibitem{Wu1}Biao Wu and Qian Niu, Phys. Rev. A \textbf{64}, 061603(R)
     (2001).

     \bibitem{Meier2}J. Meier, G. I. Stegeman, D. N. Christodoulides, Y. Silberberg, R.
     Morandotti, H. Yang, G. Salamo, M. Sorel, and J. S. Aitchison, Phys.
     Rev. Lett. \textbf{92}, 163902 (2004).

     \bibitem{Stepic}M. Stepi\'{c}, C. Wirth, C. E. R\"{u}er, and D. Kip, Opt.
     Lett. \textbf{31}, 247 (2006); C. E. R\"{u}er, J. Wisniewski, M. Stepi\'{c}, and D.
     Kip, Opt. Express \textbf{15}, 6324 (2007).

     \bibitem{Smerzi}A. Smerzi, A. Trombettoni, P. G. Kevrekidis, and A. R.
     Bishop, Phys. Rev. Lett. \textbf{89}, 170402 (2002).

     \bibitem{Machholm1}M. Machholm, C. J. Pethick, and H. Smith, Phys. Rev. A
     \textbf{67}, 053613 (2003).

     \bibitem{Burger} S. Burger, F. S. Cataliotti, C. Fort, F. Minardi, and M.
     Inguscio, M. L. Chiofalo, and M. P. Tosi, Phys.
     Rev. Lett. \textbf{86}, 4447 (2001).

     \bibitem{Fallani} L. Fallani, L. De Sarlo, J. E. Lye, M. Modugno, R. Saers, C. Fort, and M.
     Inguscio, Phys. Rev. Lett. \textbf{93}, 140406 (2004).

     \bibitem{Louis} P. J. Y. Louis, E. A. Ostrovskaya, C. M. Savage, and Y. S. Kivshar,  Phys. Rev.
     A \textbf{67}, 013602 (2003).

     \bibitem{Efremidis} N. K. Efremidis and D. N. Christodoulides, Phys. Rev.
     A  \textbf{67}, 063608 (2003).

     \bibitem{Mayteevarunyoo} T. Mayteevarunyoo and B. A. Malomed, Phys. Rev.
     A \textbf{74}, 033616 (2006).

     \bibitem{Kartashov} Y. V. Kartashov, V. A. Vysloukh, and L. Torner, Opt.
     Express \textbf{12}, 2831 (2004).

     \bibitem{Xu}Z. Xu, Y. V. Kartashov, and L. Torner, Phys. Rev. Lett.
     \textbf{95}, 113901 (2005).

     \bibitem{Sukhorukov}A. A. Sukhorukov and Y. S. Kivshar, Opt. Lett. \textbf{28}, 2345 (2003).

     \bibitem{Cohen} O. Cohen, T. Schwartz, J. W. Fleischer, M. Segev, and D. N. Christodoulides, Phys. Rev.
     Lett. \textbf{91}, 113901 (2003);  A. A. Sukhorukov  and  Y. S. Kivshar, Phys. Rev. Lett. \textbf{91}, 113902 (2003).

     \bibitem{Meier}J. Meier, J. Hudock, D. Christodoulides, G. Stegeman, Y. Silberberg, R. Morandotti, and J.
     S. Aitchison, Phys. Rev. Lett. \textbf{91}, 143907 (2003); Z. Chen, A.
     Bezryadina, I. Makasyuk, and J. Yang, Opt. Lett. \textbf{29},1656 (2004).


     \bibitem{Pelinovsky} D. E. Pelinovsky, A. A. Sukhorukov,
     and Y. S. Kivshar, Phys. Rev. E \textbf{70}, 036618 (2004).

     \bibitem{Kartashov1} Y. V. Kartashov, V. A. Vysloukh, and
     L. Torner, Phys. Rev. Lett. \textbf{96}, 073901 (2006).

     \bibitem{Eiermann} B. Eiermann, T. Anker, M. Albiez, M. Taglieber, P. Treutlein, K. P. Marzlin, and
     M. K. Oberthaler,  Phys. Rev. Lett. \textbf{92}, 230401 (2004).






    \bibitem{Yang}J. Yang and Z. Musslimani, Opt. Lett.
    \textbf{23}, 2094 (2003); E. A. Ostrovskaya and Y. S. Kivshar,
    Phys. Rev. Lett. \textbf{93}, 160405 (2004).

    \bibitem{Malomed}B. A. Malomed and P. G. Kevrekidis,  Phys. Rev.
    E \textbf{64}, 026601 (2001); M. \"{O}ster and M. Johansson,  Phys. Rev.
    E \textbf{73}, 066608 (2006); P. G. Kevrekidis, H. Susanto, and Z.
    Chen,  Phys. Rev. E  \textbf{74}, 066606 (2006).



    \bibitem{Carr} L. D. Carr, C. W. Clark, and W. P. Reinhardt, Phys. Rev. A \textbf{62},
    063610 (2000); \textbf{62}, 063611 (2000).



    \bibitem{Alexander} T. J. Alexander, E. A. Ostrovskaya, and Yuri S. Kivshar,
    Phys. Rev. Lett. \textbf{96}, 040401 (2006).

    \bibitem{Smirnov}E. Smirnov, C. E. R\"uter, D. Kip, Y. V. Kartashov,
    and L. Torner, Opt. Lett. \textbf{32}, 1950 (2007).


    \bibitem{Zhang} Yongping Zhang and Biao Wu, Phys. Rev. Lett.
    \textbf{102}, 093905 (2009).

  \bibitem{Liang} Z. X. Liang, B. B. Hu and Biao Wu, e-print arXiv:cond-mat/0903.4058.

    \bibitem{Jordan}D. W. Jordan and P. Smith, Nonlinear Ordinary Differential
    Equations (Clarendon Press, Oxford, 1977).

    \bibitem{Wu2}Biao Wu and Qian Niu, New J. Phys. \textbf{5}, 104
    (2003).

\bibitem{Wang2}Jiandong Wang, Jianke Yang, Tristram J. Alexander, and Yuri S.
   Kivshar, Phys. Rev. A \textbf{79}, 043610 (2009).

    \bibitem{Machholm} M. Machholm, A. Nicolin, C. J. Pethick, and H. Smith, Phys. Rev. A \textbf{69},
     043604 (2004).

    \bibitem{Diakonov} D. Diakonov, L. M. Jensen, C. J. Pethick, and H.
    Smith, Phys. Rev. A \textbf{66}, 013604 (2002); M. Machholm, C. J. Pethick, and H.
    Smith, Phys. Rev. A \textbf{67}, 053613 (2003).

    \bibitem{Dong} X. Dong and Biao Wu, Laser Phys. {\bf 17}, 190 (2007).

    \bibitem{Mandelik}D. Mandelik, R. Morandotti, J. S. Aitchison,
    and Y. Silberberg, Phys. Rev. Lett. \textbf{92}, 093904 (2004).

    \bibitem{Neshev}D. Neshev, A. A. Sukhorukov, B. Hanna, W. Krolikowski, and Y. S.
    Kivshar, Phys. Rev. Lett. \textbf{93}, 083905 (2004).


    \bibitem{Wang} J. Wang and J. Yang, Phys. Rev. A \textbf{77}, 033834 (2008).





\end{thebibliography}
\end{document}